\def\ts     {\thinspace}
\def\kms    {\ifmmode{{\rm \ts km\ts s}^{-1}}\else{\ts km\ts s$^{-1}$}\fi}
\def\msol   {\ifmmode{{\rm M}_{\odot} }\else{M$_{\odot}$}\fi}
\def\lsol   {\ifmmode{L_{\odot}}\else{$L_{\odot}$}\fi}
\def\lfir   {\ifmmode{L_{\rm FIR}}\else{$L_{\rm FIR}$}\fi}
\def\zsol   {\ifmmode{{\rm Z}_{\odot}}\else{Z$_{\odot}$}\fi}
\def\etal   {{\rm et\ts al.\ }}
\def\aco    {\ifmmode{{\rm CO}(J\!=\!1\! \to \!0)}\else{{\rm CO}($J$=1$\to$0)}\fi}
\def\bco    {\ifmmode{{\rm CO}(J\!=\!2\! \to \!1)}\else{{\rm CO}($J$=2$\to$1)}\fi}
\def\cco    {\ifmmode{{\rm CO}(J\!=\!3\! \to \!2)}\else{{\rm CO}($J$=3$\to$2)}\fi}
\def\dco    {\ifmmode{{\rm CO}(J\!=\!4\! \to \!3)}\else{{\rm CO}($J$=4$\to$3)}\fi}
\def\eco    {\ifmmode{{\rm CO}(J\!=\!5\! \to \!4)}\else{{\rm CO}($J$=5$\to$4)}\fi}
\def\fco    {\ifmmode{{\rm CO}(J\!=\!6\! \to \!5)}\else{{\rm CO}($J$=6$\to$5)}\fi}
\def\gco    {\ifmmode{{\rm CO}(J\!=\!7\! \to \!6)}\else{{\rm CO}($J$=7$\to$6)}\fi}
\def\hco    {\ifmmode{{\rm CO}(J\!=\!8\! \to \!7)}\else{{\rm CO}($J$=8$\to$7)}\fi}
\def\ico    {\ifmmode{{\rm CO}(J\!=\!9\! \to \!8)}\else{{\rm CO}($J$=9$\to$8)}\fi}
\def\jco    {\ifmmode{{\rm CO}(J\!=\!10\! \to \!9)}\else{{\rm CO}($J$=10$\to$9)}\fi}
\def\kco    {\ifmmode{{\rm CO}(J\!=\!11\! \to \!10)}\else{{\rm CO}($J$=11$\to$10)}\fi}
\def\ci {\mbox{\rm C~{\scriptsize I}}}
\def\hi     {\ifmmode{{\rm H}{\rm \small I}}\else{H\ts {\scriptsize I}}\fi}
\def\hh     {\ifmmode{{\rm H}_2}\else{H$_2$}\fi}
\def\cone {\ifmmode{{\rm C}{\rm \small I}(^3\!P_1\!\to^3\!P_0)}
     \else{C\ts {\scriptsize I}{\small$(^3\!P_1\!\to^3\!\!\!P_0)$}}\fi}
\def\ctwo {\ifmmode{{\rm C}{\rm \small I}(^3\!P_2\!\to^3\!P_1)}
     \else{C\ts {\scriptsize I}{\small$(^3\!P_2\!\to^3\!\!\!P_1)$}}\fi}
\def\cij {\ifmmode{{\rm C}{\rm \small I}\,(^3P_i\to^3P_j)}
\else{C\ts {\scriptsize I}\,{\small$(^3P_i\to^3P_j)$}}\fi}
\def\cii    {\ifmmode{{\rm C}{\rm \small II}}\else{C\ts {\scriptsize II}}\fi}
\def\tex {\ifmmode{{T}_{\rm ex}}\else{$T_{\rm ex}$}\fi}
\def\tmb {\ifmmode{{T}_{\rm mb}}\else{$T_{\rm mb}$}\fi}
\def\tkin {\ifmmode{{T}_{\rm kin}}\else{$T_{\rm kin}$}\fi}
\def\microns {\ifmmode{\mu{\rm m}}\else{$\mu$m}\fi}
\def\nhh   {\ifmmode{n({\rm H}_2)}\else{$n$(H$_2$)}\fi}
\def\gradv {\ifmmode{(dv/dr)}\else{$(dv/dr)$}\fi}
\def\CO10{{\hbox {CO(1$-$0)}}}

\def\,{\thinspace}

\def\msun{M$_\odot$}

\def \kkmspc{K\,\kms\,pc$^2$}
\def \ppm{$\pm$}

\documentclass[longauth]{aa}
\usepackage{graphicx}
\usepackage{epsfig}
\usepackage{txfonts}
\begin{document}

   \title{The CO line SED and atomic carbon in IRAS F10214+4724\thanks{Based on
   observations carried out with the IRAM Plateau de Bure Interferometer.
   IRAM is supported by INSU/CNRS (France), MPG (Germany) and IGN (Spain).}}
   \author{Y. Ao \inst{1,2}\thanks{email: ypao@pmo.ac.cn}, A. Wei\ss\ \inst{2}, D. Downes \inst{3}, F. Walter \inst{4}, C. Henkel \inst{2} and K. M. Menten \inst{2}}
   \institute{
   Purple Mountain Observatory, Chinese Academy of Sciences, Nanjing 210008, China
   \and
   MPIfR, Auf dem H\"{u}gel 69, 53121 Bonn, Germany
   \and
   IRAM, Domaine Universitaire, 38406 St-Martin-d'H\`{e}res, France
   \and
   MPIA, K\"{o}nigstuhl 17, 69117 Heidelberg, Germany }
   \date{}

\authorrunning{Y. Ao et al.}
\titlerunning{The CO line SED and atomic carbon in IRAS F10214+4724}


\abstract{ Using the IRAM 30m telescope and the Plateau de Bure
interferometer we have detected the \ctwo\, and the CO 3$-$2, 4$-$3, 6$-$5,
7$-$6 transitions as well as the dust continuum at 3 and
1.2\,mm towards the distant luminous infrared galaxy IRAS F10214+4724 at
$z=2.286$. The \ctwo\ line is detected for the first time towards this source and IRAS
F10214+4724 now belongs to a sample of only 3 extragalactic sources
at any redshift where both of the carbon fine structure lines have
been detected. The source is spatially resolved by our \ctwo\
observation and we detect a velocity gradient along the east-west
direction. The CI line ratio allows us to derive a carbon excitation
temperature of 42$^{+12}_{-9}$~K. The carbon excitation in
conjunction with the CO ladder and the dust continuum constrain the gas density to
$n(\hh)$\,=\,$10^{3.6-4.0}$ cm$^{-3}$ and the kinetic temperature to
$T\rm_{kin}$\,=\,45--80~K, similar to the excitation conditions
found in nearby starburst galaxies. The rest-frame 360\,$\mu$m dust
continuum morphology is more compact than the line emitting region,
which supports previous findings that the far infrared luminosity arises from
regions closer to the active galactic nucleus at the center of this system.

 }

\maketitle

\section{Introduction}
Recent submm- and mm-wavelength dust continuum surveys from SCUBA at the JCMT
(Ivison et al. 2000; Coppin et al. 2006) and MAMBO at the IRAM 30m
telescope (Bertoldi et al. 2007) have revealed a population of
so-called submm galaxies (SMGs) at high redshift with far infrared
(FIR) luminosities comparable to or higher than those of local
Ultraluminous Infrared Galaxies (ULIRGs). Spectroscopic follow-up
studies of SMGs in the optical regime suggest that the volume density of
these sources increases by three orders of magnitude out to $z\sim2$
(Chapman et al. 2005) and, in contrast to the local Universe,
ULIRGs could thus dominate the total bolometric luminosity from star
formation at early epochs.\\

\noindent About 50 of these high-z, luminous infrared (IR)
galaxies have been detected in CO emission lines (see Solomon \&
Vanden Bout 2005 for a review), and a significant fraction have
been imaged in CO with the IRAM Plateau de Bure Interferometer (PdBI)
by Greve et al. (2005). Such observations provide important constraints on
their molecular gas masses, kinematics and dynamical masses. For a
few of these sources multi-transition studies of CO, \ci\ and HCN
have been presented (Wei\ss\ et al. 2005b, 2007; Alloin et al.
2007), which focus on the excitation properties of these massive
gas reservoirs at high redshift.\\

\noindent IRAS F10214+4724 (F10214 hereafter) was the first high-z
source detected in CO (Brown \& Vanden Bout 1991; Solomon et al.
1992) and has been observed in the CO(3$-$2) line by many authors (see
Radford et al. 1996 for a summary), and the CO(4$-$3) (Brown \& Vanden Bout
1992), CO(6$-$5) (Solomon et al. 1992) as well as the \cone\, transitions (Wei\ss\
et al. 2005a). CO(1$-$0) was tentatively detected (Tsuboi \& Nakai
1992, 1994), but not confirmed by Barvainis (1995). An upper limit
on the \ctwo\ line was reported by Papadopoulos (2005). The gas distribution
was also investigated in detail by the high angular resolution observations in CO(3$-$2)
of Downes et al. (1995) and Scoville et al. (1995).\\

\noindent In this paper we present new measurements of the
CO(3$-$2), CO(4$-$3), CO(6$-$5) and CO(7$-$6) lines and the first
successful detection of the higher level fine structure line of atomic
carbon, together with the continuum emission at 3\,mm and 1.2\,mm towards F10214.


\section{Observations}
\subsection{IRAM PdBI observations}
\noindent We observed the CO $J$=3$-$2 rotational line and the \ctwo\ fine structure line with the PdBI
in the compact D configuration during 8
nights in 2004. The dual frequency setup was used, and assuming a redshift, z, of 2.2854 the receivers
were tuned to the redshifted CO(3$-$2) frequency of 105.252 GHz in the 3\,mm band
and the frequency of the redshifted \ctwo\ transition at 246.345 GHz in the 1.2\,mm band (both upper side band
tunings). During this run the weather was poor and the performance
of the old receivers in the 1.2\,mm band was unsatisfactory, thus only the 3\,mm
data was useful, yielding an equivalent 6-antenna on-source
integration time of 17~hours for CO(3$-$2) with an average system
temperature of 180\,K. Baselines ranged from 15.5 to 112.6 meters
resulting in a synthesized beam of 6.5$\arcsec\times$3.8$\arcsec$
(P.A. $\sim$ 105$\rm ^o$ E of N) for natural weighting and
4.9$\arcsec\times$3.2$\arcsec$ (P.A. $\sim$ 92$\rm ^o$ E of N) for
uniform weighting. The spectral correlator covers $\sim$1590 \kms\
with a velocity resolution of 2 \kms.\\

\noindent The \ctwo\ line was reobserved with the new PdBI receivers on December
25th 2007  under good weather conditions in C configuration. Both polarizations
were tuned to the redshifted \ctwo\ line, giving a velocity coverage of $\sim$1210 \kms\ (1GHz
band width) and a velocity resolution of 3 \kms.
Baselines ranged from 16.3 to 175.3 meters yielding a synthesized beam
of 1.12$\arcsec\times$0.97$\arcsec$ (P.A. $\sim$ 4$\rm ^o$ E of N) for natural
weighting. The on-source integration time for this run is $\sim$5.5~hours with
a typical system temperature of 200\,K.\\

\noindent Amplitudes were calibrated using 3C84, 3C454.3, 0420-014, CRL618, and MWC349.
For phase calibration we observed the nearby calibrators 0923+392 and
0955+476. The data were processed with the programs from GILDAS software packages and the final data cubes have
noise levels of 1.9 mJy per beam (for uniform weighting) at 20 \kms\
and 2.7 mJy per beam at 30 \kms\ resolution for the 3\,mm and 1.2\,mm line cubes, respectively.
The noise levels of the final continuum maps, which were computed by averaging the
line free channels, are 0.17 mJy per beam (for natural weighting) at 3\,mm and 0.7 mJy
per beam at 1.2\,mm, respectively. We estimate the flux density scale to be accurate to
about 10\% at 3\,mm and 20\% at 1.2\,mm.\\

\subsection{IRAM 30m observations}
\noindent The IRAM 30m observations toward F10214 were carried out
during the period between 2004 July and 2005 February. We used the A/B
and C/D receiver combinations tuned to CO(3$-$2) (105.252 GHz, 3\,mm
band), CO(4$-$3) (140.330 GHz, 2\,mm band), CO(6$-$5) (210.468 GHz,
1.3\,mm band) and CO(7$-$6) (245.526 GHz, 1.2\,mm band) lines,
respectively. The beam sizes/antenna gains for increasing
frequency are 23.4$\arcsec$/6.3 Jy K$^{-1}$, 17.5$\arcsec$/6.7 Jy
K$^{-1}$, 11.7$\arcsec$/7.9 Jy K$^{-1}$ and 10$\arcsec$/9.5 Jy
K$^{-1}$, respectively. Typical system temperatures were about 120
K, 255 K, 240 K and 330 K for the four CO lines, respectively. The
observations were done in wobbler switching mode, with a switching
frequency of 0.5 Hz and a wobbler throw of 50$\arcsec$ in azimuth.
Data were recorded using the 512\,$\times$\,1 MHz filter banks
for the 3\,mm receivers (500 MHz total bandwidth), and the
256\,$\times$\,4 MHz filter banks for the 2 and 1.3\,mm receivers
(1 GHz total bandwidth). The data were reduced with the CLASS
software.  We first omitted scans with distorted baselines, and
only subtracted linear baselines from individual spectra, and then
averaged all the useful scans. Finally, we smoothed the spectra
to the velocity resolution of 34, 34, 40 and 30 \kms,
respectively. The useful on-source integration times are 3.8, 5.9,
3.2 and 9.2 hours for the four CO lines and the resulting rms noise
values are 3.5, 4.0, 3.3 and 4.6 mJy, respectively.



\section{Results}

\subsection{Neutral carbon line}\label{carbon-result}

\noindent The \ctwo\ line is detected towards F10214 for the first
time. The integrated line flux is $I_\ctwo$\,=\,4.59\ppm0.66 Jy \kms
which is consistent with the upper limit of 7 Jy \kms\ reported by
Papadopoulos (2005). Together with the detection of the \cone\ line
(Wei\ss\ et al. 2005a), F10214 is only the third extragalactic
source, next to the Cloverleaf (Barvainis et al. 1997; Wei\ss\ et
al. 2003) and M82 (Stutzki et al. 1997), where both carbon lines
have been detected. The resulting carbon line ratio, $\rm
$L$^{\prime}_{\ctwo}/$L$^{\prime}_{\cone}$, is 0.84\ppm0.19. The \ctwo\
line width is $182\pm30$\kms\ which agrees well with the linewidth
derived from the CO(3$-$2) and CO(6$-$5) lines and the lower carbon
line (see $\S~\ref{co-lines}$).\\

\begin{figure*}
  \centering
   \includegraphics[angle=-90,width=0.75\textwidth]{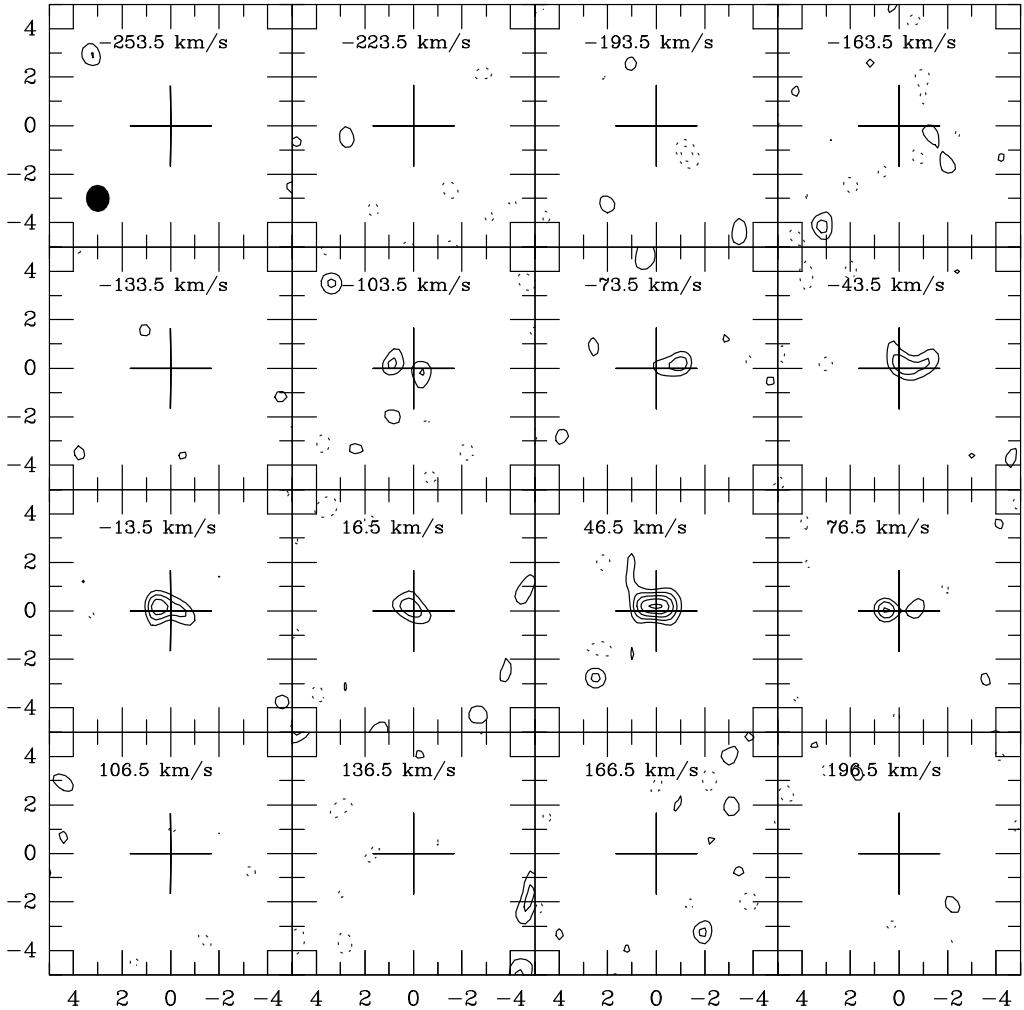}
   \caption{Channel maps of the \ctwo\ line of F10214 after subtracting the continuum.
   The contours are -3, -2, 2, 3, 4,
   5, 6, 7 $\times$ 2.7 mJy~beam$^{-1}$ (1~$\sigma$), with a synthesized beam of
1.12$\arcsec$$\times$0.97$\arcsec$, which is shown in the lower left corner
of the first panel. The maps are centered on the position
shown as a cross ($\alpha$(J2000)\,=\,10$\rm^h$24$\rm^m$34$\rm^s$.56, $\delta$(J2000)\,=\,+47$\rm
^o$09$\arcmin$09$\arcsec$.8), here and in subsequent figures.}\label{f10214-1mm-line}
\end{figure*}

\begin{figure*}
  \centering
   \includegraphics[angle=-90,width=0.5\textwidth]{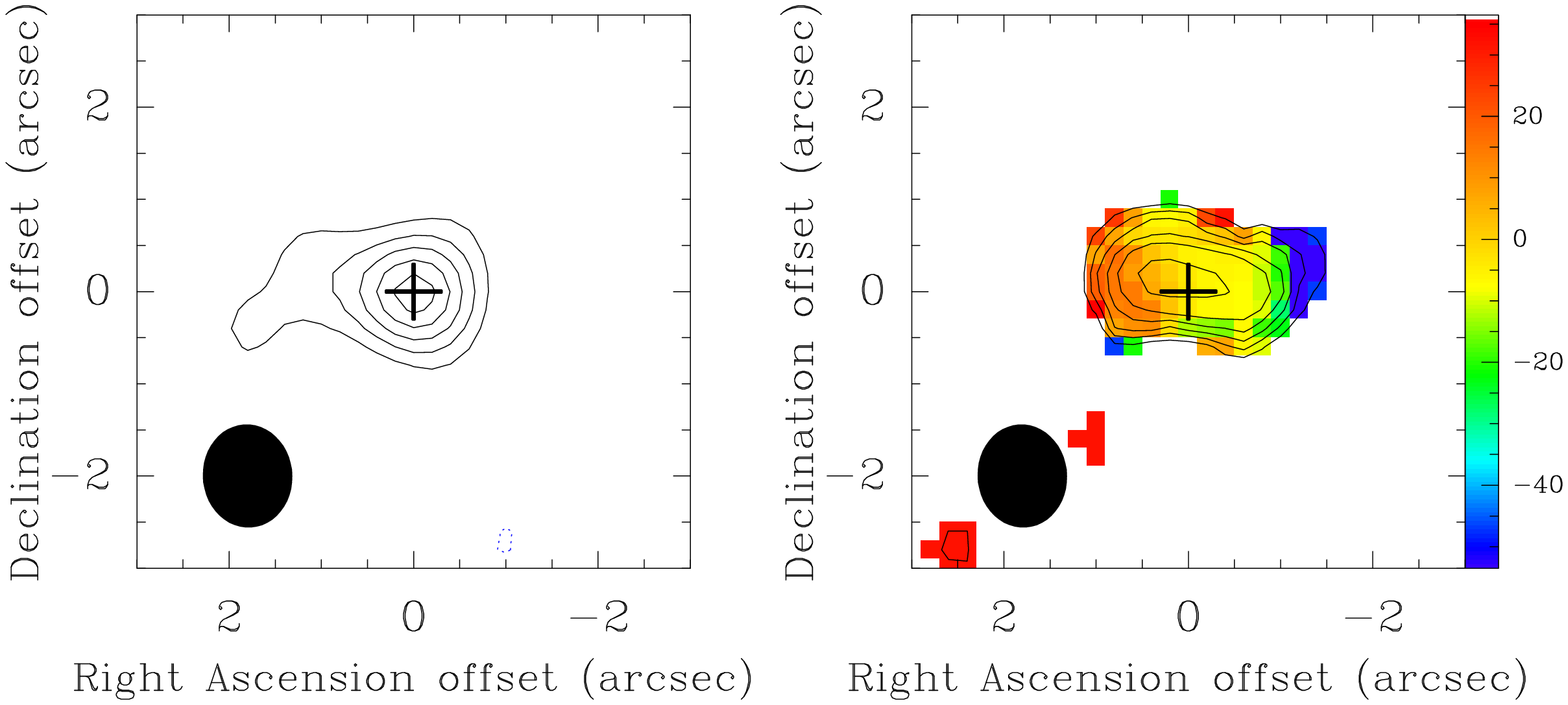}
   \caption{Left panel: Contour map of the 1.2mm continuum towards F10214,
   which was created from the line free channels. Contours are -3, 3, 5, 7, 9,
   11 $\times$ 0.7 mJy beam$^{-1}$ (1~$\sigma$).
   Right panel: Contour map of the \ctwo\ line integrated intensity towards F10214
   after subtracting the continuum overlaid by the velocity field in color map.
   Contours are -3, 3, 6, 9, 12, 15, 20
   $\times$ 0.081 Jy~\kms\ beam$^{-1}$ (1~$\sigma$).
   A synthesized beam of 1.12$\arcsec$$\times$0.97$\arcsec$ is
   shown in the lower left corner of the figures.}\label{f10214-1mm-map}
\end{figure*}

\noindent The \ctwo\ spectral line data are presented as channel maps
in Fig.~\ref{f10214-1mm-line}. The emission peaks move from west to east
for increasing velocities which indicates that the gas rotates along the east-west
direction. This rotation is also seen in the velocity field of the \ctwo\ line,
which is presented as a color map in Fig.~\ref{f10214-1mm-map} (right). In the same figure we
show the contours of the integrated line intensity. From the line intensity
map it is apparent that the \ctwo\ distribution is spatially resolved.
The deconvolved full width to half maximum (FWHM) source size is $<0.7\arcsec$ and 1.4$\arcsec$$\pm$0.1$\arcsec$
along the minor and major axis respectively, and the position angle is 84$\rm ^o$ (E of N).
This morphology is similar to the arc-like lensed structure visible at
2.2\microns\ (Matthews \etal\ 1994, Graham \& Liu 1995).
The carbon distribution, however, does not show the additional compact 2.2\microns\
component about 3$\arcsec$ north of the arc which argues for differential lensing
between the wavelengths of the near infrared (NIR) emission and those of the molecular gas. The \ctwo\ distribution is similar to the
morphology derived from CO (Downes \etal\ 1995).\\

\noindent The 1.2\,mm (restframe 360 $\mu$m) continuum map is shown
in Fig~\ref{f10214-1mm-map} (left). Interestingly, its distribution
differs from the morphology seen in the carbon line. It it much more
compact and the deconvolved FWHM source size is $<$0.7$\arcsec$. The continuum peaks at
$\alpha(\rm J2000)$\,=\,10$\rm^h$24$\rm^m$34$\rm^s$.56, $\delta(\rm
J2000)$\,=\,+47$\rm^o$09$\arcmin$09$\arcsec$.8 ($\pm0.1''$) which agrees
well with the peak of the \ctwo\ line. The 1.2\,mm continuum flux is
9.9\ppm1.2 mJy, in good agreement with the flux of 9.6\ppm1.4 mJy
reported by Downes et al. (1992) at the same wavelength. The peak
intensity is 8.3 mJy~beam$^{-1}$, which suggests the source is
slightly resolved spatially.

\subsection{CO lines}\label{co-lines}

\begin{figure*}
  \centering
   \includegraphics[angle=-90,width=0.75\textwidth]{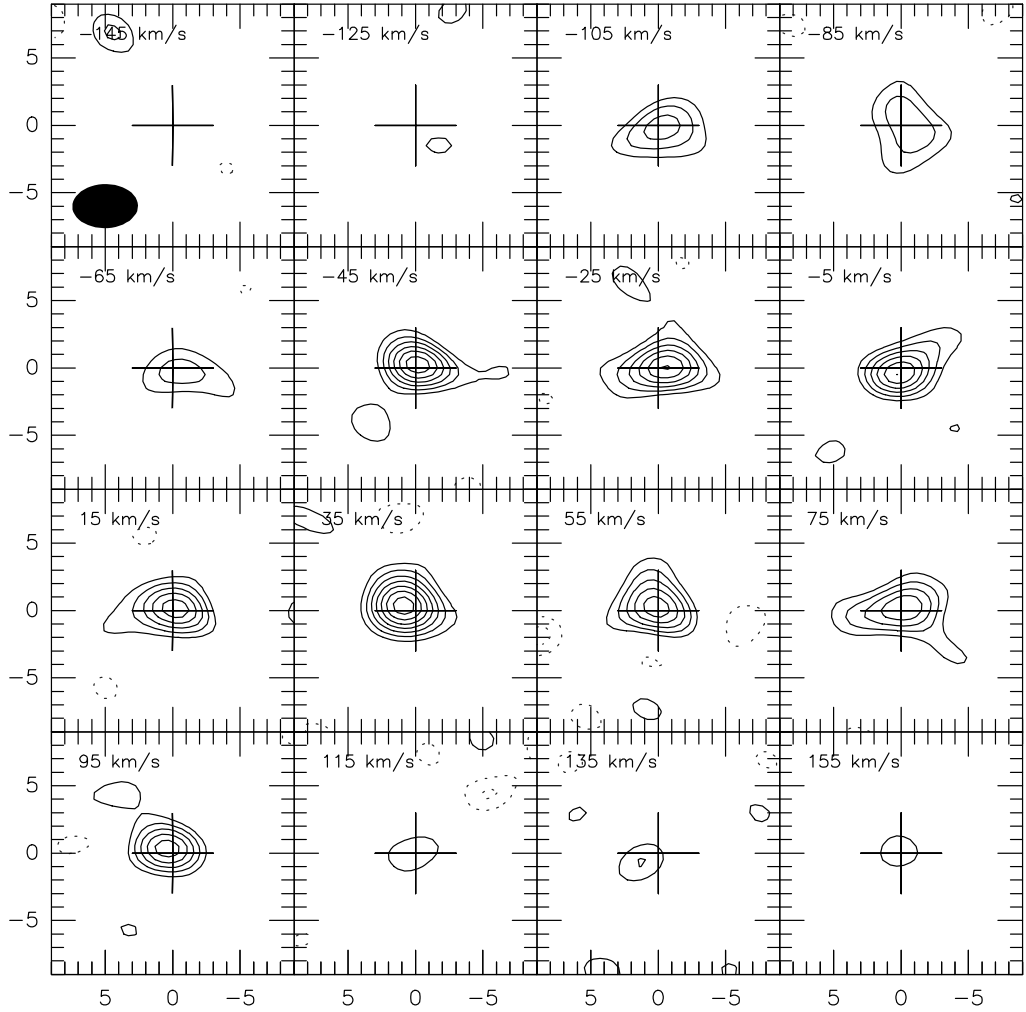}
   \caption{CO(3$-$2) channel maps of F10214 after subtracting the continuum.
   The contours are -3, -2, 2, 3, 4, 5, 6, 7, 8, 9 $\times$ 1.9 mJy~beam$^{-1}$
   (1~$\sigma$), with a synthesized beam of 4.9$\arcsec$$\times$3.3$\arcsec$,
   which is shown in the lower left corner of the first panel.}\label{f10214-3mm-line}
\end{figure*}

\begin{figure*}
  \centering
   \includegraphics[angle=-90,width=0.5\textwidth]{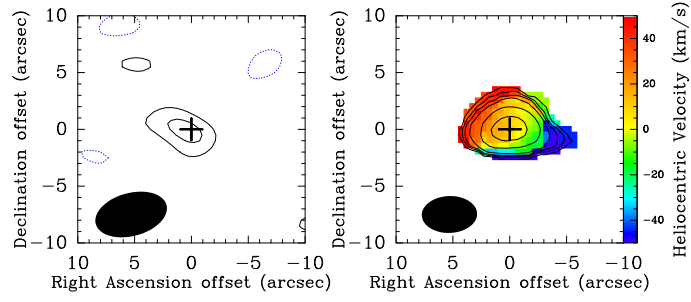}
   \caption{Left panel: Contour map of the 3mm continuum towards F10214, which
   was created from the line free channels. Contours are -2, 2, 3 $\times$
   0.17 mJy beam$^{-1}$ (1~$\sigma$).
   Right panel: Contour map of the CO(3$-$2) line integrated intensity towards F10214
   after subtracting the continuum overlaid by the velocity field in color map.
   Contours are -3, 3, 9, 15, 30, 50 $\times$
   0.038 Jy~\kms\ beam$^{-1}$ (1~$\sigma$). The synthesized beams of
   6.5$\arcsec$$\times$3.8$\arcsec$ and 4.9$\arcsec$$\times$3.3$\arcsec$ are
   shown in the lower left corner of the figures.}\label{f10214-3mm-map}
\end{figure*}

\noindent Our PdBI CO(3$-$2) line observations are represented as channel
maps in Fig.~\ref{f10214-3mm-line} and the moment maps in the right
panel of Fig.~\ref{f10214-3mm-map}. The 3\,mm continuum
(rest frame 930\microns) is detected for the first time and we find
an integrated flux density of 0.58\ppm0.24 mJy and an emission peak at a
3$\sigma$ level. To achieve higher sensitivity, natural
weighting was used to create the continuum map
(Fig.~\ref{f10214-3mm-map}\,left). Although the signal is only
detected at 3$\sigma$ significance, the good positional agreement
with the 1.2\,mm peak supports the reliability of our detection. The 3mm continuum as
well as the integrated line distribution remain unresolved in our
4.9$\arcsec$$\times$3.3$\arcsec$ and
6.5$\arcsec$$\times$3.8$\arcsec$ beams, consistent with the size
estimates from the higher resolution 1.2\,mm data. The emission
centroids of the integrated CO(3$-$2) and the continuum distributions
coincide well with the peak position derived at 1.2\,mm. Inspection
of individual channel maps shows that also the CO(3$-$2) emission
has a velocity gradient along east-west direction, consistent with
the rotation seen in \ctwo. The CO(3$-$2) velocity field is shown in
Fig.~\ref{f10214-3mm-map}~(right), where the ordered velocity gradient is obvious
along the east-west direction. The spatial separation between the
centroids of the blue and red emission in the channel maps is about
$2\arcsec$, somewhat higher than, but consistent with, the extent of
the \ctwo\ emission. This further suggests that the molecular gas
detected in CO is cospatial with the gas detected in the carbon
line.\\

\begin{figure*}
  \centering
   \includegraphics[angle=-90,width=1\textwidth]{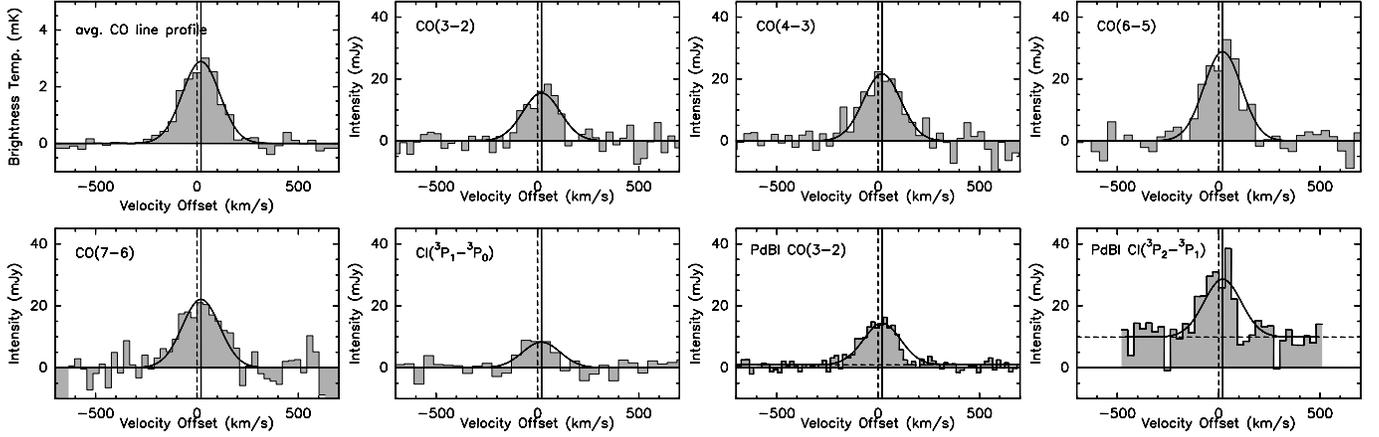}
   \caption{Spectra of the averaged profile and the CO 3$-$2, 4$-$3, 6$-$5,
   7$-$6, \cone\ and \ctwo\ lines towards F10214. The results from the PdBI are
   shown in the two panels at the lower right. All lines are fitted and Gaussian fits
   are shown as continuous lines for which the FWHM line width and the central velocity are fixed to
   the value determined from the averaged CO profile.
   The dashed horizontal lines in the PdBI spectra show the dust continuum levels.
   The zero and Gaussian fitted velocities from the averaged CO profile are
   shown as dashed vertical lines. The \cone\ spectrum is taken
   from Wei\ss\ et al. (2005a).}\label{f10214-spectrum}
\end{figure*}

\noindent Our 30m CO spectra are presented in Fig.~\ref{f10214-spectrum}. To reduce the
uncertainty in the CO line profiles we use all CO spectra to obtain
an average line profile (first panel of Fig.~\ref{f10214-spectrum}). A
Gaussian fit to this spectrum yields a line width of 246\ppm10 \kms,
and redshift of 2.28562\ppm0.00004. Line parameters of individual
lines were obtained from Gaussian fits keeping the linewidth fixed
to the value of the average spectrum. The line parameters are
summarized in Table \ref{f10214-table}. The averaged CO line width is higher
than the values from both carbon lines and the CO(3$-$2) line, which is
mainly because the CO(4$-$3) and CO(7$-$6) appear to have broader line profiles.
which is mainly because the CO(4-3) and CO(7-6) lines appear to have
broader profiles. More specifically both CO(3-2) spectra are somewhat
asymmetric with the blue line wing being more prominent than the red
wing (see Fig.~\ref{f10214-spectrum}). This asymmetry is also marginally visible in both 
CI profiles but not in the higher-J CO transitions. Although the effect
is not very pronounced it could indicate the presence of cold foreground
material at low density with sufficient optical depth to affect the low-J 
CO and CI lines but not the high J CO transitions which have much higher
critical densities. One could also argue that only the low-J CO and CI 
lines arise cospatially while the high-J CO transitions trace a 
different volume. The later explanation, however, ignores that a volume
that emits in the high-J CO transitions will always be bright in the 
low-J lines too.\\

\noindent The CO(3$-$2) flux of the spectrum from the 30m telescope
is 3.80\ppm0.45 Jy \kms, which is in agreement with previous
results (see Radford et al. 1996 for a summary) and also
with our high S/N PdBI observation which gives
3.40\ppm0.19 Jy \kms.\\

\noindent Previous observations of the CO(4$-$3) and CO(6$-$5)
lines have been reported by Brown \& Vanden Bout (1992) and Solomon et
al. (1992). Our CO(4$-$3) integrated intensity is
5.32$\pm$0.51 Jy \kms, which is only about one third of the value given
by Brown \& Vanden Bout (1992). Considering the high quality of
our new spectrum and the flux densities observed for the other CO
transitions we consider our new CO(4$-$3) measurement to be more
reliable. Our CO(6$-$5) line integrated intensity is 7.09$\pm$0.47
Jy \kms, which is consistent with the previous value by Solomon et
al. (1992).\\

\begin{figure*}
  \centering
   \includegraphics[angle=0,width=0.5\textwidth]{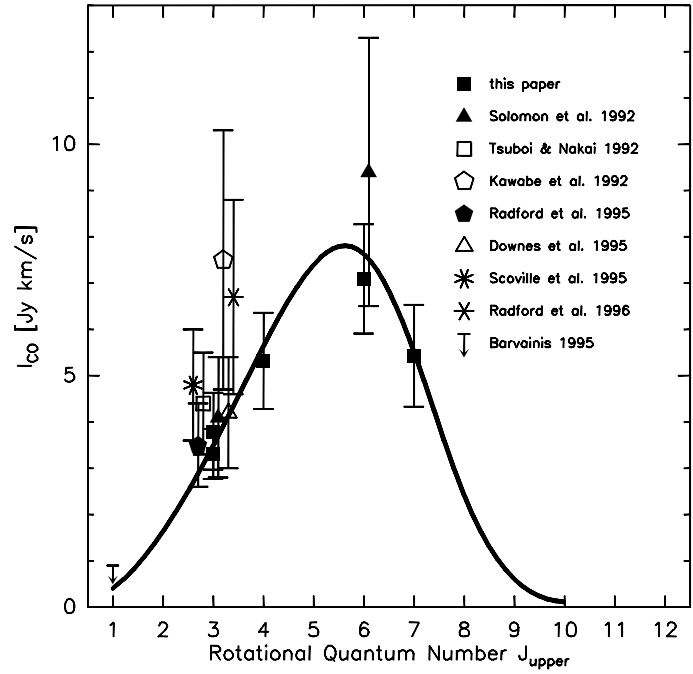}
   \caption{Observed CO fluxes vs. rotational quantum number (CO line SED,
   filled squares) for F10214 fitted by the a single component LVG model with
   $n$(H$_2$)\,=\,10$^{3.8}$ cm$^{-3}$ and $T_{\rm kin}$\,=\,60 K, shown as a
   thick solid line. Other measurements, taken from the literatures, are marked
   by different symbols. For better visibility, previous data points for
   the 3$-$2 line are shown with small x-axis offsets.}\label{f10214-single}
\end{figure*}

\noindent Our detection of the CO(7$-$6) transition is the first published
detection of this transition towards this source.
Its line integrated intensity is 5.43$\pm$0.56 Jy \kms. Therefore,
our observations constrain the peak of the CO line spectral
energy distribution (SED; i.e. the integrated line flux densities versus the
rotational quantum number) in F10214 to be at the CO(6$-$5) line. The CO line SED
including most of the CO observations from the literature and the present study is shown in
Fig.~\ref{f10214-single}.

\section{Discussion}

\subsection{CO gas excitation}\label{colvg}

\begin{figure*}
  \centering
   \includegraphics[angle=0,width=0.5\textwidth]{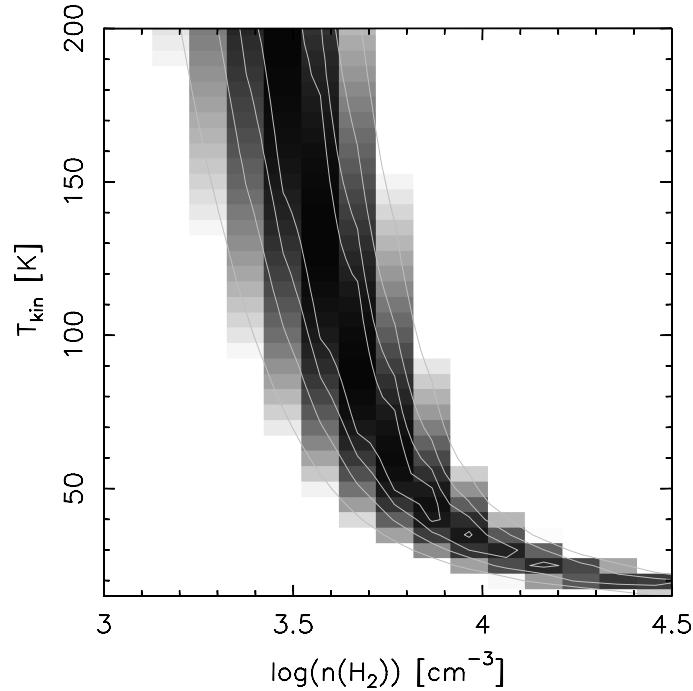}
   \caption{Reduced $\chi ^2$ distribution for a single component LVG model fit to the
   observed line luminosity ratios (grey scale and white contours, contours:
   $\chi ^2\,=\,$0.5, 1, 2, 4). The CO abundance per velocity gradient for the
   LVG models is [CO]/ $\mbox{({\rm d}v/{\rm d}r) }$\,=\, $1\times 10^{-5}~{\rm pc}~
   (\thinspace \rm km\thinspace \rm s^{-1})^{-1}$.}\label{f10214-chi-lvg}
\end{figure*}

To investigate the CO excitation we here use a one-component large velocity gradient (LVG) analysis as
described by Wei\ss\ et al. (2007) adopting a fixed CO abundance per
velocity gradient of [CO]/$\rm{(dv/dr)}$\,=\,1$\times$ 10$^{\rm -5}$
pc\,($\rm \kms$)$^{-1}$.
As discussed before, the line profiles may suggest a more complex gas excitation 
(for an example of a two component model, see the case of APM08279+5255
discussed by Wei\ss\ et al. 2007). Since, however, the
differences in the line profiles are not very pronounced and more
likely explained by foreground absorption rather than a 2 component
excitation we use here for simplicity a one-component model which
yields the average gas excitation in F10214.
From the LVG model we can limit the allowed
range for the H$_2$ density to 10$^{3.3-4.2}$ cm$^{-3}$. The
kinetic temperature of the gas, however, is poorly constrained.
This is shown in Fig.~\ref{f10214-chi-lvg} where we show the $\chi ^2$
distribution of the $T\rm _{kin}$--n(H$\rm _2$) parameter space of
the CO LVG models.\\

\noindent For a gas kinetic temperature of $T\rm_{kin}$\,=\,45--80\,K
(as suggested from our carbon analysis and the dust model, see below)
the allowed range for the gas density narrows to n(\hh)\,=\,$10^{3.6-4.0}$ cm$^{-3}$.
For these gas parameters the low-J CO lines are optically thick and thermalized
which implies $L^{\prime}\rm_{CO(1-0)}\approx$\,$L^{\prime}\rm_{CO(3-2)}$.
The LVG predicted integrated intensity of the CO(1$-$0) and CO(2$-$1) line emission are
0.37\ppm0.02 Jy \kms\ and 1.55\ppm0.05 Jy \kms, respectively.
Within the uncertainties, the CO(1$-$0) flux from the LVG models is
in agreement with the upper limit derived by Barvainis (1995).

\subsection{Excitation temperature from \ci\ }

Recent studies in the Milky Way and nearby galaxies have shown that
CO and \ci\ have very similar distributions supporting the
interpretation that their emission arises from the same gas volume
(Fixsen et al. 1999; Ojha et al. 2001; Ikeda et al. 2002). The
results from the carbon line and the CO(3$-$2) line by the PdBI also
support that the molecular gas detected in CO is cospatial with the
gas detected in the carbon line, as mentioned in $\S$3. Thus, we can
use our observed \ci\ line ratio to obtain an independent estimate
of the kinetic temperature via the \ci\ excitation temperature to
solve the ambiguity between the kinetic temperature and density
arising from the CO LVG models. As discussed in Schneider et al.
(2003) the carbon excitation temperature in the local thermodynamic
equilibrium (LTE) can be derived from the \ci\ line ratio via the
formula $T_{\rm ex}$\,=\,$\frac{38.8}{ln(\frac{2.11}{\rm R})}$,
where \small $R$\,=\,$\int T_{mb}(\ctwo)dv/\int T_{mb}(\cone)dv$,
\normalsize assuming that both carbon lines share the same
excitation temperature and are optically thin. With our observed
line ratio of 0.84\ppm0.19, we find $T\rm _{ex}$\,=\,
42$^{+12}_{-9}$ K.


\subsection{\ci\ abundance estimates}
\noindent Using our estimate of the CO(1$-$0) line luminosity and the standard
ULIRG factor $X$$\rm_{CO}$\,=\,0.8 M$\rm_\odot$ $\rm(K\ \kms\ pc^2)^{-1}$
(Downes \& Solomon 1998), we find a molecular gas mass of
(8.9$\pm$1.0)$\times$10$^{10}$\,m$^{-1}$\,M$\rm_\odot$ where m is the
magnification factor by the gravitational lens. Using Eq.\,2 in Wei\ss\ et al.
(2005a) and our carbon excitation temperature of 42\,K, the neutral
carbon mass is $M_{\ci}$\,=\,(3.7\ppm0.7)$\times$10$^7$ m$^{-1}$ \msun.
This leads to a carbon abundance, [\ci]/[H$_2$]\,=\,$M$[\ci]/6$M$[H$_2$], of
(6.9\ppm1.2)$\times\,10^{-5}$, in agreement with our previous results (Wei\ss\
et al. 2005a).


\subsection{\ci\ LVG models}\label{ci-lvg}

To obtain an independent test on the optical depth of the \ci\ lines
in F10214 as well as on the LTE assumption used to derive the
excitation temperature, we have also calculated LVG models for \ci\
(Stutzki et al. 1997) using $\mbox{[\ci]/\rm{(dv/dr)}}\,=\,1.4\times
10^{\rm -5}$ pc\,($\rm \kms$)$^{-1}$. This abundance per velocity gradient corresponds to
the \ci\ abundance estimated above and a velocity gradient of
$\mbox{({\rm d}v/{\rm d}r)}\,=\,5 \rm \kms\ pc^{-1}$. For \tkin\,=\,60\,K
and $n$(\hh)\,=\,$10^{3.8}$cm$^{-3}$, which provides a good fit to the
observed CO SED, the resulting \ctwo/\cone\ line ratio is 0.8, which is
consistent with our measurement of 0.84\ppm0.19. The
LVG model predicts a \cone/CO(3$-$2) line ratio of 0.38, which is
somewhat higher than the observed ratio of 0.26\ppm0.05. For this
solution the excitation temperatures for the lower and upper carbon
lines are 54 and 45\,K, respectively. The optical depth in both
transitions ranges from 0.5 to 0.6. Fine tuning of the gas parameters (e.g.
a somewhat higher kinetic temperature of 70--80\,K or a slightly
lower carbon abundance per velocity gradient) will also bring the
\cone/CO(3$-$2) line ratio in agreement with the observations without
significant changes of the excitation temperature or the optical
depth of the lines. This implies that CO and CI LVG models with
kinetic temperatures of $\sim45-80$\,K provide reasonable
estimates of the physical gas properties.
This further suggests that the carbon lines in F10214
are indeed not far from LTE. Given the underlying assumptions,
the single component LVG model for CO and CI gives a reasonable
prediction for all CO and \ci\ line intensities. This finding
supports the view that the \ci\ and CO emission arises from the same
volume on galactic scales.

\subsection{Size and magnification for CO and \ci}

The measured size of the \ci\ emitting region in conjunction with
the intrinsic line brightness temperatures for CO and \ci\ allow us
to estimate the lensing magnification factor and the intrinsic size of the
gas distribution (see Downes \etal 1995). From our LVG
models for \ci\ and CO we find an intrinsic brightness temperature
of 10--14\,K and 38--47\,K for \ctwo\ and CO(3$-$2) respectively. Assuming
our measured arc-length of $1.4\arcsec$ also holds for the CO lines and
adopting the method in Downes \etal (1995) to derive the lens magnification factor,
this yields magnifications of 11--16 for the carbon line and 9--12 for CO(3$-$2),
in agreement with the results in Downes \etal (1995). We note that
the different magnifications between both lines simply reflect the
uncertainties of the underlying brightness temperature and do not
imply differential magnification. For a magnification factor of $m$\,=\,12, the
mean between both estimates, we derive an intrinsic source radius of
$\sim 490$\,pc.

\subsection{The dust continuum}\label{dust-constrains}

\noindent We have included our new PdBI 3\,mm and 1.2\,mm
measurements in the dust continuum analysis using a 2-component dust
model as described in Wei\ss\ et al. (2007).
Because the optically thin approximation for the dust emission
does not necessarily hold at rest wavelength shorter than $\sim$100 $\mu$m 
for ULIRGs (Downes et al. 1993) we have used the approach described
in Wei\ss\ et al 2007: 
the flux density of the dust emission is related to the dust temperature,
$\rm T_{dust}$, and the apparent solid angle, $\Omega _{app}$ by
the relation $\rm
S_{\nu}=[B_{\nu}(T_{dust})-B_{\nu}(T_{bg})]\frac{(1-e^{-\tau _{\nu}})}{(1+z)^3}\Omega _{app}$,
where $\rm S_{\nu}$ is the observed (amplified) flux density and
$\rm B_{\nu}$ is the Planck function. The lens magnification factor is
hidden in the apparent solid angle, $\Omega _{app}$. The dust optical depth, $\tau _{\nu}$,
is a property of the source itself and is independent of the effects of gravitational lensing:
$\tau _{\nu}=\kappa_d(\nu _r)M_{dust,app}/D_A^2 \Omega _{app}$
where $\rm D_A$ is the angular distance of the source, $\rm M_{dust,app}$
is the apparent dust mass and $\rm \kappa_d$ is the dust absorption
coefficient. For the dust absorption coefficient, we adopt $\rm
\kappa_d(\nu) = 0.4~(\nu _r/250~GHz)^\beta$ cm$^2$ g$^{-1}$, where
the dust emissivity index, $\beta$=2.\\

\begin{figure*}
  \centering
   \includegraphics[angle=0,width=0.5\textwidth]{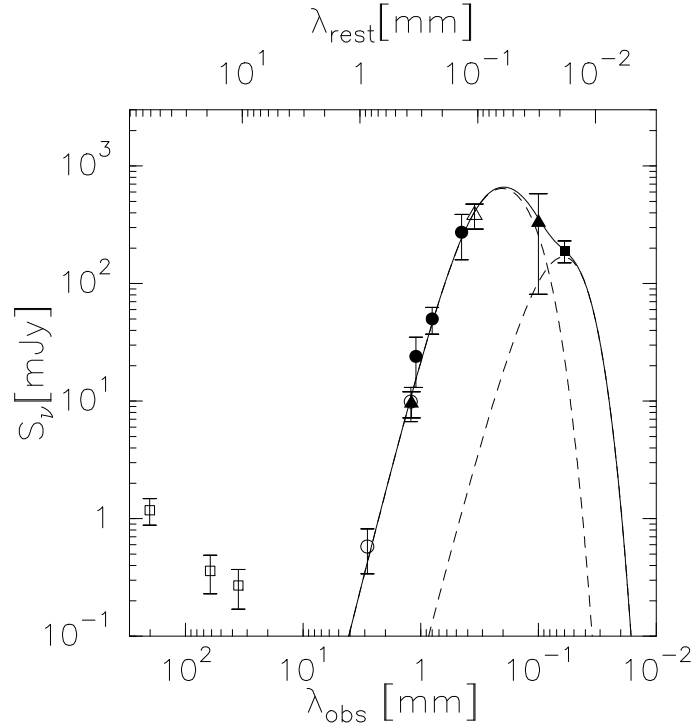}
   \caption{Two component dust model
   for F10214. Displayed flux densities were taken from Lawrence et al. 1993
   (20~cm, 6~cm and 3.6~cm, open squares), this work (3\,mm and 1.2\,mm, open circles),
   Downes et al. 1992 (1200, 100 $\mu$m, filled triangles),
   Rowan-Robinson et al. 1993 (1100, 800, 450 $\mu$m, filled circles),
   Benford et al. 1999 (350 $\mu$m, open triangle) and
   Rowan-Robinson et al. 1991 (60 $\mu$m, filled square).
   The dashed lines show the thermal dust continuum
   emission for 80 and 160 K dust components. The solid line is the total
   emission from both components. See $\S~\ref{dust-constrains}$ for the details
   of the dust models.}\label{f10214-dust-model}
\end{figure*}

\noindent The fit to all the data between 3\,mm and 60 $\mu$m is shown
in Fig.~\ref{f10214-dust-model}. The contribution of the non-thermal
radio emission is negligible even at 3\,mm.  For the ``cold''
component, we find a dust temperature, $T\rm_{cold}$, of
80\ppm10 K and a dust mass, M(cold dust), of 1.1$\times$10$^9$
m$^{-1}$ M$\rm _\odot$(\ppm25\%). The warm component is poorly
characterized as only two data points are available for the relevant frequency
range. Our dust temperature for the cold component is in
good agreement with the value given in Downes \etal\ (1992), but
higher than the value of 55\,K reported by Benford \etal\ (1999)
although they adopted a slightly lower value of the dust emissivity
index of 1.5. The physical size of the radius implied by our
dust model is $>350$\,pc for a magnification factor of 12. 
Much smaller real radii are inconsistent with the observed 
slope of the Rayleigh-Jeans tail of
the dust SED due to the increasing optical depth of the dust
emission. For the ``cold'' component, the dust optical depth
at 100 $\mu$m of the rest frame is about 1.9. The dust temperature
for the ``cold'' component provides an additional constraint on the
kinetic temperature, which helps to narrow the ambiguity of the CO
LVG models.\\

\noindent The small observed source size ($<0.7\arcsec$, i.e. $<5800$\,pc,
leads to a physical size of the radius of $<240$\,pc for a magnification factor of 12)
of the dust continuum emission at 1.2\,mm implies
that the FIR emission in F10214 arises from a different region than
the \ci\ and CO lines. This conclusion has already been reached by
Downes \etal\ (1995) based on surface brightness arguments. The
degree of compactness, however, is surprising because it is
difficult to explain this structure in the context of existing lens
models. Eisenhardt \etal (1996) suggested a
magnification for the FIR emission region of 30 which translates
into an even smaller FIR radius ($<100$\,pc). Such small regions are
inconsistent with our dust SED models unless we would adopt
a dust emissivity spectral index $\beta > 2$
to compensate the higher optical depth arising from the more
compact source size. We note, however, that our 1.2\,mm dust
continuum image shows a very faint east-west extension visible in
the lowest contour shown in Fig.~\ref{f10214-1mm-map}\,(left). We
therefore cannot rule out that we underestimate the size of the
continuum region in our data. More sensitive 1.2\,mm observations at
ideally higher spatial resolution will be required to settle this
issue. On the other hand, our high spatial resolution observations
recover all of the 1.2\,mm flux detected in single dish bolometer
observations (Downes \etal\ 1992), so it appears unlikely that our
observations miss large amounts of more extended emission. Therefore,
the FIR emitting region may be compact enough that AGN heating could
play a significant role even for the cooler dust at 80\,K. This is
also supported by the fact that the cold dust in F10214 is warmer
than 40-60\,K typically found in high-z QSOs (Benford \etal
1999; Beelen \etal\ 2006).

\subsection{Comparison to other galaxies}

\begin{figure*}
  \centering
   \includegraphics[angle=0,width=0.5\textwidth]{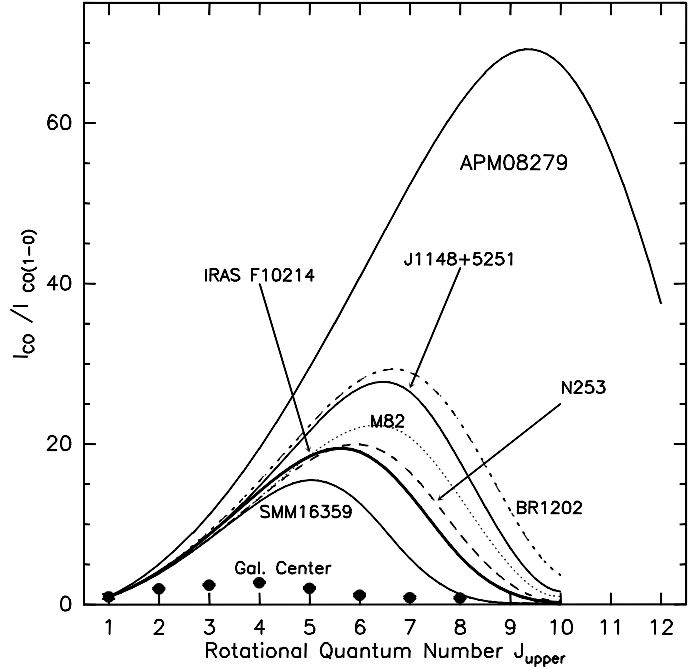}
   \caption{Comparison of the CO line SEDs of selected local and high-z
   galaxies. The SEDs are shown for IRAS F10214+4724 (thick solid line,
   as shown in Fig.~\ref{f10214-single}), APM\,08279+5255 (z\,=\,3.9, Wei\ss\ et al. 2007),
   BR\,1202-0725 (z\,=\,4.7, Riechers et al. 2006), SDSS J1148+5251 (z\,=\,6.4,
   Bertoldi et al. 2003; Walter et al. 2003), the high-excitation component
   in the center of M82 (Wei\ss\ et al. 2005c), NGC\,253 center
   (G\"{u}sten et al. 2006), SMM J16359+6612 (z\,=\,2.5, Wei\ss\ et al. 2005b)
   and the Galactic Center (solid circles, Fixsen et al. 1999). The CO line
   SEDs are normalized by their CO(1$-$0) flux density.}\label{f10214-comparison}
\end{figure*}

Only a limited number of sources have been observed in a large
enough  number of CO lines to allow for a determination of the peak of their CO line
SEDs (Fig.~\ref{f10214-comparison}). From this figure it is apparent that
F10214 has CO excitation characteristics similar to the centers of the nearby
galaxies NGC\,253 and M82, and high-z galaxies like J1148+5251
and BR1202-0725 (for references, see the caption to
Fig.~\ref{f10214-comparison}). For all these sources, the peak of the CO
SED occurs at the CO(6$-$5) or CO(7$-$6) lines. It is interesting to note, that for local
starbursts such as M82 the nuclear region is surrounded by large
amounts of gas with lower CO excitation. This gas even dominates
the emission in CO transitions below the CO(5$-$4) line (Walter et
al. 2002; Wei\ss\ et al. 2005c). No such low excitation
component is apparent in the CO SEDs of the high-z sources studied
so far (Riechers et al. 2006). This implies that \textit{all} the molecular gas and, 
in consequence, the starbursts
arise from compact regions that dominate the (sub)mm emission.
This is in agreement with the small equivalent radii determined
from the dust and CO modelling, and also with the gas morphology
of nearby ULIRGs where the CO emission often arises from compact,
bright nuclear regions (Downes \& Solomon 1998). So far, high-J CO
observations in local ULIRGs have only been reported for Mrk231
where the highest observed line is the CO(6$-$5) transition (Papadopoulos
et al. 2007). Also for this source the CO SED rises up to the
CO(6$-$5) transition, which implies similar or even higher gas excitation compared
to high-z FIR luminous galaxies.\\

\noindent While the CO excitation appears to be very similar for
most active galaxies, there are two outstanding examples visible in
Fig.~\ref{f10214-comparison}: APM08279+5255 has a much higher CO
excitation compared to all other sources. For this source, AGN
heating and high gas densities have been suggested to be the main
driver of the high gas excitation (Wei\ss\ et al. 2007). The submm
galaxy SMM16359+6612, on the other hand, has the lowest CO
excitation in the sample. For this source it was suggested that it
is a merger that is not yet in the typical ULIRG stage as most of
the gas may still be located in the overlap region between the
merging galaxies (Kneib et al. 2004; Wei\ss\ et al. 2005b). For such
a geometry the average gas densities (and therefore also the CO
excitation as measured by the peak of the CO SED) is expected to be
lower than for typical ULIRGs where almost all gas is located in the
very central region of the merged galaxies.

\section{Conclusion}
Using the IRAM 30m telescope and the PdBI we have detected the
\ctwo, CO $J$=3$-$2, 4$-$3, 6$-$5 and 7$-$6 transitions as well as the
dust continuum at 3 and 1.2\,mm towards IRAS F10214+4724.
The \ctwo\ line is detected for the first
time towards this source and F10214 is now one out of only three
extragalactic objects at any redshift where both carbon fine
structure lines have been detected. The CI line ratio allows us to
derive a carbon excitation temperature of 42$^{+12}_{-9}$~K.\\

\noindent We have used the \ci\ excitation temperature to narrow the
temperature-density ambiguity in the LVG modeling for CO. The CO and
\ci\ lines together with the dust continuum constrain the gas density
to $n$(\hh)\,=\,$10^{3.6-4.0}$ cm$^{-3}$ and the gas kinetic temperature to
$T\rm_{kin}$\,=\,45--80~K. The gas excitation in F10214 is therefore
similar to that found in the nuclei of nearby starburst galaxies.
Given the underlying assumptions, the single component LVG model for
CO and CI gives a reasonable prediction for all CO and \ci\ line
intensities. This finding supports the view that the \ci\ and CO
emission arises from the same volume on galactic scales.\\

\noindent The source is well resolved by our new \ctwo\ line
observation along the east-west direction, showing extended emission
and a velocity gradient along the major axis,
which is similar to the structure seen in CO. The major
axis diameter in conjunction with the intrinsic brightness
temperatures of CO and \ci\ imply a gravitational magnification of
$\sim12$, in agreement with previous results.\\

\noindent The continuum emission at 1.2\,mm is more compact than the
morphology of the \ctwo\ line which shows that the FIR emitting
region is smaller than the molecular gas distribution. We find a
temperature of the cold dust of $T\rm_{cold}$\,=\,80\ppm10 K. The high
dust temperature together with the compact morphology suggests that
parts of the FIR emission could be due to heating of the central
AGN.


\begin{acknowledgements}
We thank P. Cox for supporting this project through DDT observing
time allocation for the new PdBI receivers which greatly improved
our \ctwo\ line spectrum.
Y.A. acknowledges the financial support from Chinese Academy of
Science for supporting his stay as a visiting scholar at MPIfR, when 
most of this work was done. Y.A. also acknowledges the support 
from NSFC grant 10733030. Finally, we appreciate the comments 
of the anonymous referee which improved our manuscript. 

\end{acknowledgements}

\begin{center}
\begin{table*}[t]
\caption{Observed CO and $\mbox{\rm C {\scriptsize I}}$ line
parameters towards F10214}\label{f10214-table} \vspace{0.5cm}
\begin{tabular}{c c c c c c c c c}
\hline Line & Telescope & $\nu_{\rm obs}$ & S$_\nu$$^{\mathrm{a}}$
& $\Delta$V$_{\rm FWHM}$$^{\mathrm{b}}$ & I$^{\mathrm{a}}$ &
V$^{\mathrm{b}}$ & $L^\prime$/10$^{\rm 10}$$^{\mathrm{\, c}}$ \\
& & [GHz] & [mJy] & [\kms] & [Jy \kms] & [\kms] & [\kkmspc] \\
\hline
avg. profile  & IRAM 30m & ... & ... & 246$\pm$10 & ... & 20$\pm$4 & ... \\
CO(3$-$2)  & IRAM 30m & 105.25230 & 15.5$\pm$3.5 & 174$\pm$23 & 3.80$\pm$0.45 & 25$\pm$11 & 10.83$\pm$1.28 \\
CO(3$-$2)  & PdBI & 105.25160  & 13.5$\pm$0.6 & 224$\pm$12 & 3.40$\pm$0.19 & 4$\pm$4 & 9.70$\pm$0.54 \\
CO(4$-$3)  & IRAM 30m & 140.33020 & 21.7$\pm$4.0 & 245$\pm$28 & 5.32$\pm$0.51 & 17$\pm$11 & 8.53$\pm$0.82 \\
CO(6$-$5)  & IRAM 30m & 210.46850 & 28.8$\pm$3.3 & 203$\pm$16 & 7.09$\pm$0.47 & 20$\pm$7 & 5.06$\pm$0.34 \\
CO(7$-$6)  & IRAM 30m & 245.52621 & 22.1$\pm$4.6 & 248$\pm$30 & 5.43$\pm$0.56 & 18$\pm$14 & 2.84$\pm$0.29 \\
\cone & IRAM 30m & 149.80235 & 8.3$\pm$2.4  & 156$\pm$26 & 2.03$\pm$0.36 & -5$\pm$14 & 2.86$\pm$0.51$^{\mathrm{d}}$ \\
\ctwo & PdBI & 246.34500  & 18.7$\pm$2.3 & 182$\pm$30 & 4.59$\pm$0.66 & -6$\pm$10 & 2.39$\pm$0.34 \\
\hline
\end{tabular}
\begin{list}{}{}
\item{$^{\mathrm{a}}$The values are derived from Gaussian
fits with the fixed line width and center velocity of the
averaged CO line profile.}
\item{$^{\mathrm{b}}$ The values are obtained from
Gaussian fits to each spectrum. The velocity offsets are
centered at a redshift of 2.2854 (Downes et al. 1995).}

\item{$^{\mathrm{c}}$ We use a $\Lambda$ cosmology with $\rm
H_{\rm0}$\,=\,73~$\rm \kms\ Mpc^{-1}$, $\rm \Omega_\Lambda$\,=\,0.74
and $\rm \Omega_{\rm m}$\,=\,0.26 (Spergel et al. 2007).}

\item{$^{\mathrm{d}}$ The data are from Wei\ss\ et al. (2005a), but
fitted with the fixed line width and central velocity.}
\end{list}
\end{table*}
\end{center}

\end{document}